\documentclass[12pt,a4paper]{article}
\usepackage{graphics}
\usepackage{epsf,epsfig}
\def\xb{\overline{x}}

\begin{document}
\textwidth=135mm
 \textheight=200mm
\begin{center}
{\bfseries Polarized and transversity GPDs in kaon
leptoproduction}
\vskip 5mm S.V. Goloskokov \vskip 5mm {\small {\it Bogoliubov
Laboratory of Theoretical  Physics,  Joint Institute for Nuclear
Research, 141980 Dubna, Russia}}
\\
\end{center}
\vskip 5mm \centerline{\bf Abstract}

We study the kaon leptoproduction  on the basis of the handbag
approach. We consider the leading-twist contribution together with
the transversity twist-3 effects which were found to be important
in the description of pseudo-scalar meson production. We present
our predictions for the cross section and spin asymmetries in the
kaon leptoproduction. \vskip 10mm

In this report, we analyze  the process of kaons leptoproduction
at large photon virtualities within the handbag approach, where
the amplitudes factorize \cite{fact} into hard subprocesses and
GPDs which keep the soft physics. Different applications of GPDs
were discussed at this conference \cite{kroll}. At the
leading-twist accuracy the reactions of kaon production are
sensitive only to the GPDs $\widetilde{H}$ and $\widetilde{E}$
which contribute to the amplitudes for longitudinally polarized
virtual photons \cite{gk09}.  It was observed that to be
consistent with experimental data on the pion leptoproduction the
contributions of transversity GPDs $H_T$ and $\bar E_T$  are
needed \cite{gk11}. Within the handbag approach the transversity
GPDs are accompanied by the twist-3 meson wave function.

We consider here the transversity $H_T$ and $\bar E_T$ effects in
the leptoproduction of kaons. We present the model results for the
cross section of the $K^+ \Lambda$ and $K^+ \Sigma^0$
leptoproduction \cite{gk11} and predictions for the spin asymmetry
in these reactions. It is shown that the $H_T$ effects are
essential in the $K^+ \Lambda$ channel while in the $K^+ \Sigma^0$
leptoproduction the $\bar E_T$ contribution is mostly important.


In what follows, we calculate  the meson leptoproduction on the
basis of the handbag approach.  The hard subprocess amplitudes are
calculated within the modified perturbative approach
\cite{sterman} in which the quark transverse degrees of freedom as
well as gluonic radiation, condensed in a Sudakov factor, are
taken into account.

The proton  non- flip or helicity-flip  amplitudes for
longitudinally polarized photons ${\cal M}^{K}_{0\pm,0+}$ can be
written in the form:
\begin{equation}\label{pip}
 {\cal M}^{K}_{0+,0+} \propto
                             [P^{K}_{0+,0+}+\langle \tilde{H}^{K}\rangle
  \rangle];\; {\cal M}^{K}_{0-,0+} \propto
\frac{\sqrt{-t^\prime}}{(m+M_{N^f})}\, [ P^{K}_{0-,0+} + \xi
\langle \widetilde{E}^{K}_{n.p.}\rangle ].
\end{equation}
The amplitudes (\ref{pip}) dominate at large $Q^2$. The
corresponding amplitudes with transversally polarized photons are
suppressed as $1/Q$.

The $P^{K}$ terms in (\ref{pip}) represent a kaon pole which
appears in this reaction for charged kaon production. We use the
kaon-barion coupling constants \cite{gk11}
\begin{equation}\label{cons}
g_{K^+ p \Lambda} \sim -13.3;\;g_{K^+ p \Sigma^0} \sim -3.5,
\end{equation}
which are close to  SU(3) predictions.

The second terms in (\ref{pip}) accumulate the handbag
contribution to the kaons production amplitude. The $<\tilde{F}>$
in (\ref{pip}) is a convolution of GPD $\tilde F$ with the hard
subprocess amplitude ${\cal H}_{0\lambda,0\lambda}(\xb,...)$:
\begin{equation}\label{ff}
<\tilde{F}>= \sum_\lambda\int_{-1}^1 d\xb
   {\cal H}_{0\lambda,0\lambda}(\xb,...) \tilde{F}(\xb,\xi,t).
\end{equation}

The  proton- hyperon transition GPDs in (\ref{ff}) can be related
with the proton GPDs by using the SU(3) flavor symmetry
\cite{frankfurt99}
\begin{equation}\label{tgpd}
  F(p \to \Lambda) \sim [2
F^u-F^d-F^s];\;  F(p \to \Sigma^0) \sim [F^d-F^s].
\end{equation}

It was found that the asymptotically dominant leading-twist
contributions are not sufficient to describe the experimental
results on leptoproduction of pseudoscalar mesons \cite{gk11}. The
data require also the contributions from the transversity GPDs.

We estimate  this contribution to the ${\cal M}_{0\pm,++}$
amplitudes by the transversity GPDs $H_T$, $\bar E_T$, which are
considered together with the twist-3 meson wave function
\cite{gk11} in the hard subprocess amplitude
\begin{equation}\label{ht}
{\cal M}^{K,tw3}_{0-,\mu+} \propto
                            \int_{-1}^1 d\xb
   {\cal H}_{0-,\mu+}(\xb,...)\,H^{K}_T,\,
{\cal M}^{K,tw3}_{0-,\mu+} \propto \frac{-t'}{4\,m}
                            \int_{-1}^1 d\xb
   {\cal H}_{0-,\mu+}(\xb,...)\,\bar E^{K}_T.
\end{equation}
The $H_T$ GPD is connected with transversity PDFs  as
\begin{equation}
  H^a_T(x,0,0)= \delta^a(x);\;\;\; \mbox{and}\;\;\;
\delta^a(x)=C\,N^a_T\, x^{1/2}\, (1-x)\,[q_a(x)+\Delta q_a(x)].
\end{equation}
We parameterize the PDF $\delta$ using the model \cite{ans}. The
double distribution representation \cite{dd} is used to calculate
GPD $H_T$. Due to different signs of $H_T^u$ and $H_T^d$ we find a
quite large $H_T$ contribution $K^+ \Lambda$ and much smaller
effect in the $K^+ \Sigma^0$ production (\ref{tgpd}).

The information on $\bar E_T$  is available only from the lattice
QCD estimations \cite{lat}. It was found that  $\bar E_T^u$ and
$\bar E_T^d$ should to be quite large, have the same sign and a
similar size. From (\ref{tgpd}) we can conclude that the $\bar
E_T$ contributions to different kaon production channels should be
similar.
\begin{figure}[h!]
\begin{center}
\begin{tabular}{cc}
\includegraphics[width=6.3cm,height=5.2cm]{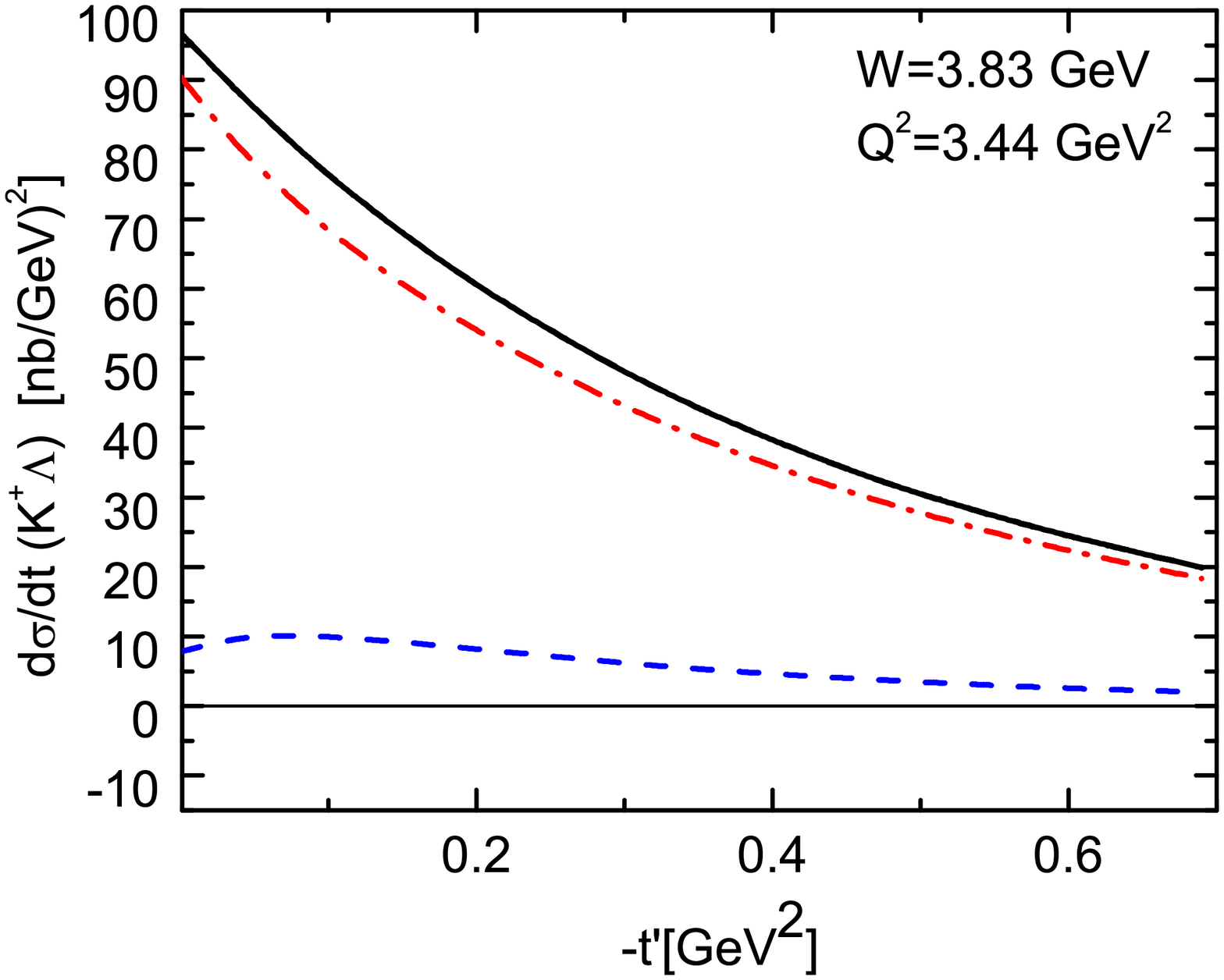}&
\includegraphics[width=6.3cm,height=5.2cm]{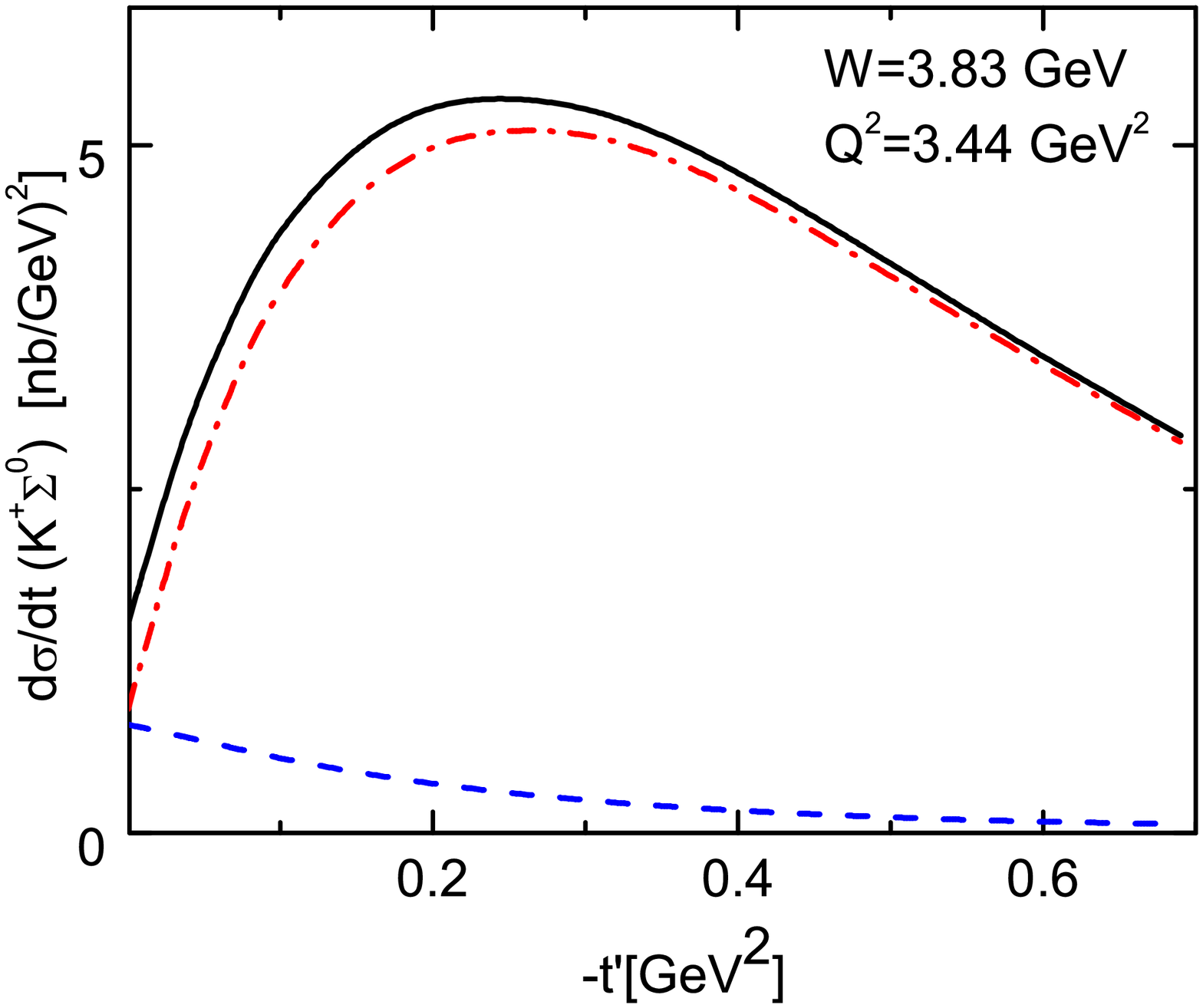}
\end{tabular}
\label{fig:5}
\end{center}
\caption{Left: the $K^+ \Lambda$ production cross sections. Right:
the $K^+ \Sigma^0$ production cross sections at HERMES energies.
Full line- unseparated cross section. dashed- $\sigma_L$,
dashed-dotted line- $\sigma_T$.}
\end{figure}

\begin{figure}[h!]
\begin{center}
\begin{tabular}{cc}
\includegraphics[width=6.3cm,height=5.2cm]{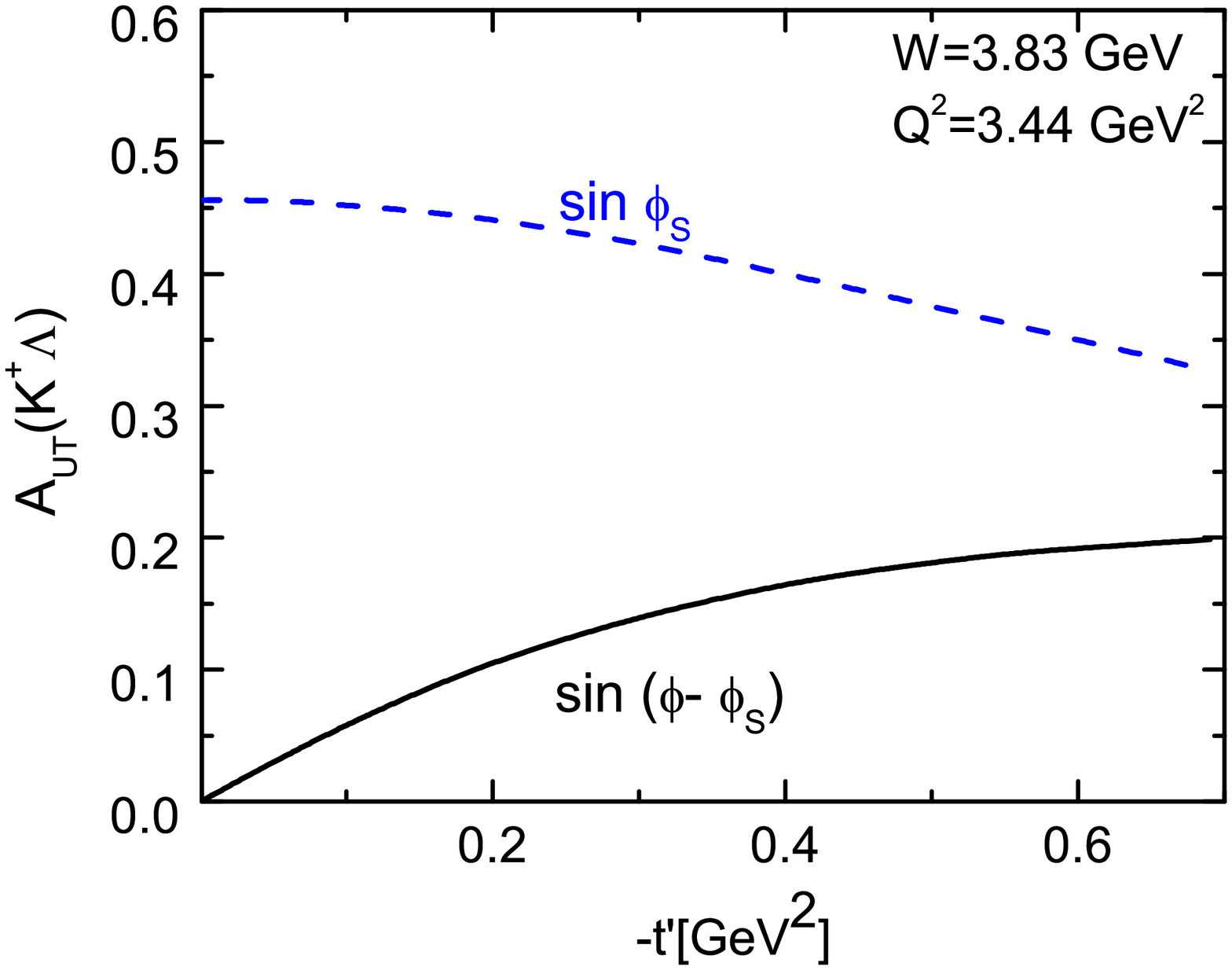}&
\includegraphics[width=6.3cm,height=5.2cm]{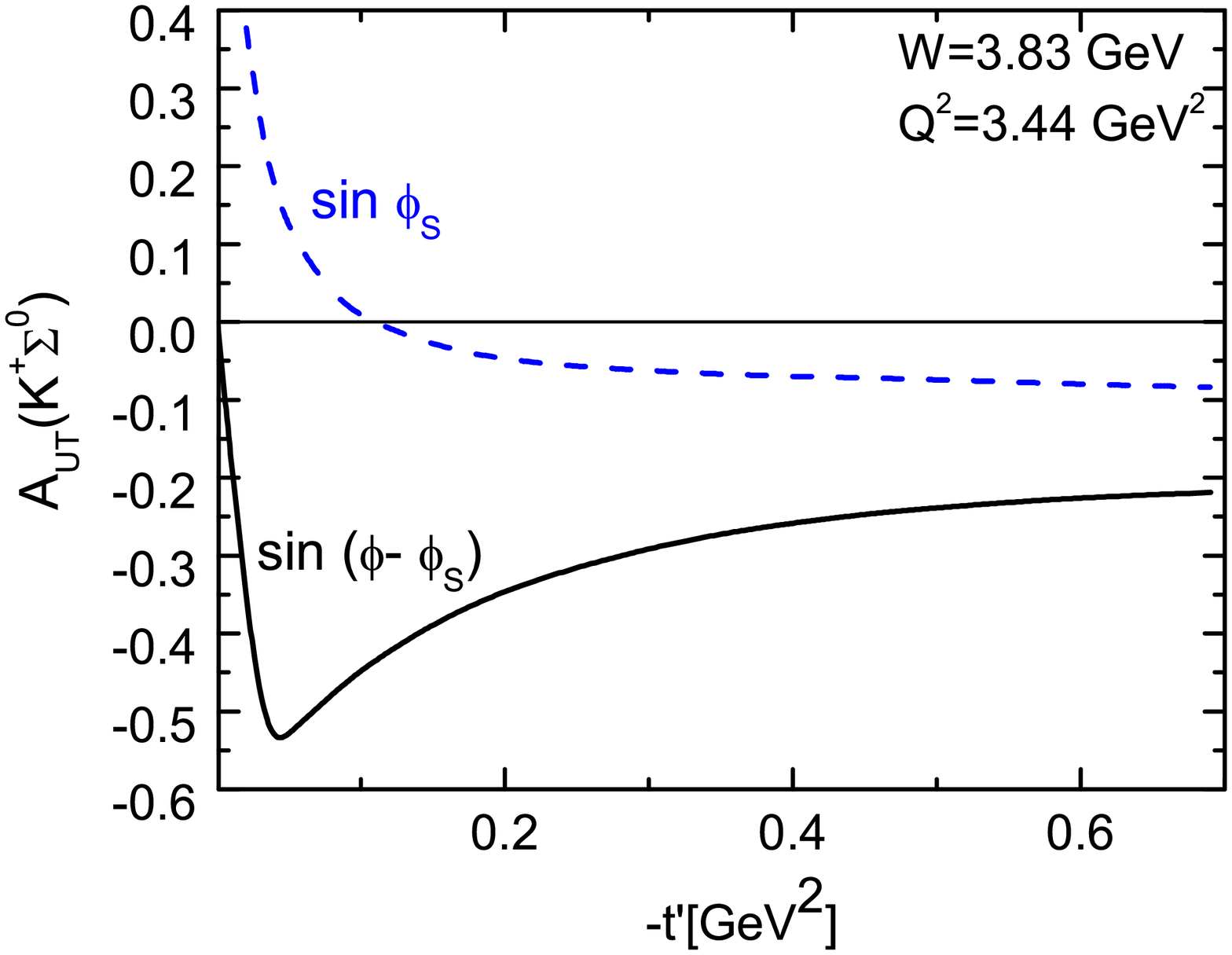}
\end{tabular}
\label{fig:5}
\end{center}
\caption{  Predicted moments of $A_{UT}$ asymmetries  at HERMES.
Left: for the $K^+ \Lambda$ channel.  Right: for the $K^+
\Sigma^0$ production.}
\end{figure}

 The large transversity $H_T$ effects
in the $K^+ \Lambda$ channel provide to the large $\sigma_T$ cross
section without a forward dip which dominated with respect to
$\sigma_L$, see Fig. 1 (Left). For the $K^+ \Sigma^0$ production
the $H_T$ contribution is much smaller and the $\bar E_T$ effects
become essential. It provides the cross section with a forward
dip, Fig. 1 (Right). In both cases $\sigma_T$ determined by the
transversity  $H_T$ and $\bar E_T$ contribution is large at low
$Q^2$ with respect to the leading twist $\sigma_L$ cross section.
Note that the twist-3 effects decrease rapidly with $Q^2$ growing
and at sufficiently high $Q^2$ the  $\sigma_L$ will predominate.

In  Fig. 2, we show our predictions for the moments of $A_{UT}$
asymmetry for kaon production. The $\sin(\phi_s)$ moment of
asymmetry determined mainly by the $H_T$ contribution is quite
large in the $K^+ \Lambda$ production, Fig. 2 (Left). In the $K^+
\Sigma^0$ channel this moment of $A_{UT}$ asymmetry is much
smaller, Fig. 2 (Right). The $\sin(\phi-\phi_s)$ moment of
$A_{UT}$ asymmetry is predicted to be not small in this process,
Fig. 2 (Right) with respect to the $K^+ \Lambda$ production.

To summarize, in this report we considered kaon leptoproduction
within the handbag approach. We calculated the leading twist and
twist-3 transversity contributions together. It was found that the
$H_T$ and $\bar E_T$ contribution was quite large. They produce
 $\sigma_T$  which at low $Q^2$ exceeds substantially the
leading twist $\sigma_L$ cross section. We observe the same effect
for most reactions of the pseudoscalar meson leptoproduction
\cite{gk11}. The role of transversity effects can be investigated
in future COMPASS and JLAB12 experiments.

This work is supported  in part by the Russian Foundation for
Basic Research, Grant  12-02-00613  and by the Heisenberg-Landau
program.

\end{document}